\documentclass[twocolumn,showpacs,preprintnumbers,amsmath,amssymb,prc,superscriptaddress]{revtex4}
\usepackage{graphicx}
\usepackage{dcolumn}
\usepackage{bm}

\begin{document}

\title{Rapidity and species dependence of particle 
production at large transverse momentum for $d$+Au 
collisions at $\sqrt{s_{\mathrm {NN}}}$ = 200 GeV}

\bigskip

\affiliation{Argonne National Laboratory, Argonne, Illinois 60439}
\affiliation{University of Birmingham, Birmingham, United Kingdom}
\affiliation{Brookhaven National Laboratory, Upton, New York 11973}
\affiliation{California Institute of Technology, Pasadena, California 91125}
\affiliation{University of California, Berkeley, California 94720}
\affiliation{University of California, Davis, California 95616}
\affiliation{University of California, Los Angeles, California 90095}
\affiliation{Carnegie Mellon University, Pittsburgh, Pennsylvania 15213}
\affiliation{University of Illinois, Chicago}
\affiliation{Creighton University, Omaha, Nebraska 68178}
\affiliation{Nuclear Physics Institute AS CR, 250 68 \v{R}e\v{z}/Prague, Czech Republic}
\affiliation{Laboratory for High Energy (JINR), Dubna, Russia}
\affiliation{Particle Physics Laboratory (JINR), Dubna, Russia}
\affiliation{University of Frankfurt, Frankfurt, Germany}
\affiliation{Institute of Physics, Bhubaneswar 751005, India}
\affiliation{Indian Institute of Technology, Mumbai, India}
\affiliation{Indiana University, Bloomington, Indiana 47408}
\affiliation{Institut de Recherches Subatomiques, Strasbourg, France}
\affiliation{University of Jammu, Jammu 180001, India}
\affiliation{Kent State University, Kent, Ohio 44242}
\affiliation{Institute of Modern Physics, Lanzhou, China}
\affiliation{Lawrence Berkeley National Laboratory, Berkeley, California 94720}
\affiliation{Massachusetts Institute of Technology, Cambridge, MA 02139-4307}
\affiliation{Max-Planck-Institut f\"ur Physik, Munich, Germany}
\affiliation{Michigan State University, East Lansing, Michigan 48824}
\affiliation{Moscow Engineering Physics Institute, Moscow Russia}
\affiliation{City College of New York, New York City, New York 10031}
\affiliation{NIKHEF and Utrecht University, Amsterdam, The Netherlands}
\affiliation{Ohio State University, Columbus, Ohio 43210}
\affiliation{Panjab University, Chandigarh 160014, India}
\affiliation{Pennsylvania State University, University Park, Pennsylvania 16802}
\affiliation{Institute of High Energy Physics, Protvino, Russia}
\affiliation{Purdue University, West Lafayette, Indiana 47907}
\affiliation{Pusan National University, Pusan, Republic of Korea}
\affiliation{University of Rajasthan, Jaipur 302004, India}
\affiliation{Rice University, Houston, Texas 77251}
\affiliation{Universidade de Sao Paulo, Sao Paulo, Brazil}
\affiliation{University of Science \& Technology of China, Hefei 230026, China}
\affiliation{Shanghai Institute of Applied Physics, Shanghai 201800, China}
\affiliation{SUBATECH, Nantes, France}
\affiliation{Texas A\&M University, College Station, Texas 77843}
\affiliation{University of Texas, Austin, Texas 78712}
\affiliation{Tsinghua University, Beijing 100084, China}
\affiliation{Valparaiso University, Valparaiso, Indiana 46383}
\affiliation{Variable Energy Cyclotron Centre, Kolkata 700064, India}
\affiliation{Warsaw University of Technology, Warsaw, Poland}
\affiliation{University of Washington, Seattle, Washington 98195}
\affiliation{Wayne State University, Detroit, Michigan 48201}
\affiliation{Institute of Particle Physics, CCNU (HZNU), Wuhan 430079, China}
\affiliation{Yale University, New Haven, Connecticut 06520}
\affiliation{University of Zagreb, Zagreb, HR-10002, Croatia}

\medskip

\author{B.I.~Abelev}\affiliation{Yale University, New Haven, Connecticut 06520}
\author{J.~Adams}\affiliation{University of Birmingham, Birmingham, United Kingdom}
\author{M.M.~Aggarwal}\affiliation{Panjab University, Chandigarh 160014, India}
\author{Z.~Ahammed}\affiliation{Variable Energy Cyclotron Centre, Kolkata 700064, India}
\author{J.~Amonett}\affiliation{Kent State University, Kent, Ohio 44242}
\author{B.D.~Anderson}\affiliation{Kent State University, Kent, Ohio 44242}
\author{M.~Anderson}\affiliation{University of California, Davis, California 95616}
\author{D.~Arkhipkin}\affiliation{Particle Physics Laboratory (JINR), Dubna, Russia}
\author{G.S.~Averichev}\affiliation{Laboratory for High Energy (JINR), Dubna, Russia}
\author{Y.~Bai}\affiliation{NIKHEF and Utrecht University, Amsterdam, The Netherlands}
% ...

\medskip
\collaboration{STAR Collaboration}\noaffiliation

\date{\today}

\bigskip

\begin{abstract}
We determine rapidity asymmetry in the production of 
charged pions, protons and anti-protons for large transverse 
momentum ($p_{\mathrm T}$) for $d$+Au collisions at 
$\sqrt{s_{\mathrm {NN}}}$~=~200~GeV. The identified hadrons 
are measured in the rapidity regions $\mid$$y$$\mid$~$<$~0.5 
and 0.5~$<$~$\mid$$y$$\mid$~$<$~1.0 for the $p_{\mathrm T}$ range 
2.5~$<$~$p_{\mathrm T}$~$<$~10~GeV/$c$.
We observe  significant rapidity asymmetry for charged pion 
and proton+anti-proton production in both rapidity regions. 
The asymmetry is larger for 0.5~$<$~$\mid$$y$$\mid$~$<$~1.0 
than for $\mid$$y$$\mid$~$<$~0.5 and is almost independent of particle type.
The measurements are compared to various model predictions employing multiple 
scattering, energy loss, nuclear shadowing, saturation effects, and
recombination, and also to a phenomenological parton model. We find that 
asymmetries are sensitive to model parameters 
and show model-preference. 
The  rapidity dependence of $\pi^{-}$/$\pi^{+}$ and $\bar{p}$/$p$ ratios
in peripheral $d$+Au and forward 
neutron-tagged events are used to study the contributions 
of valence quarks and gluons to particle production at 
high $p_{\mathrm T}$. The results are compared to calculations based on NLO pQCD 
and other measurements of quark fragmentation functions. 
\end{abstract}

\maketitle

\section{INTRODUCTION}

The mechanisms for particle production in $d$+Au collisions at RHIC may be 
different at forward and backward rapidities. The partons from the
deuteron-side (forward rapidity) are expected to undergo multiple 
scattering while traversing the gold nucleus. Those on the gold-side 
(backward rapidity) are likely to be affected by 
the properties of the nucleus. A comparative study of particle production
at forward 
and backward 
rapidity can be carried out using a ratio called the 
rapidity asymmetry ($Y_{\mathrm{Asym}}$), which is 
defined as,
\begin{displaymath}
\nonumber
Y_{\rm{Asym}}(p_{\rm T})\,
=\,\frac{Y_{\mathrm B}(p_{\mathrm T})}{Y_{\mathrm F}(p_{\mathrm T})},
\end{displaymath}
where $Y_{\mathrm F}$ and $Y_{\mathrm B}$ are forward and backward particle yields, respectively.
$Y_{\mathrm{Asym}}$ may provide unique information to help determine the 
relative contributions of various physics processes to particle
production, such as multiple scattering, nuclear shadowing, 
recombination of thermal partons, and parton saturation.

Recently, models incorporating different 
physics effects have described the nuclear modification factor for $d$+Au 
collisions ($R_{\mathrm {dAu}}$). Models including shadowing 
effects or nuclear modifications to the nucleon parton distributions 
reproduce reasonably well $R_{\mathrm {dAu}}$ for inclusive charged 
hadrons~\cite{vogt}. Those based on transverse momentum broadening 
(Cronin effect~\cite{cronin}), dynamical 
shadowing, and energy loss in cold nuclear matter~\cite{vitev}, 
also give $R_{\mathrm {dAu}}$ predictions for  inclusive charged 
hadrons, consistent with experimental data. 
Models based on the color glass condensate (CGC) approach reproduce the 
$p_{\mathrm T}$ dependence of inclusive 
charged hadron $R_{\mathrm {dAu}}$ at both mid- and forward-rapidity~\cite{cgc}. 
These models also qualitatively describe the 
pseudorapidity asymmetry for inclusive charged hadrons in 
$d$+Au collisions~\cite{star_asym}.

Another approach based on hadronization by recombination of 
thermal partons at lower $p_{\mathrm T}$ has been quite successful 
in describing the observed $R_{\mathrm {dAu}}$ for charged hadrons 
at RHIC~\cite{hwa}. This approach emphasizes  the 
hadronization portion of the final state interaction. Although it 
takes into account the hard scattering in pQCD, the fragmentation 
is replaced by recombination of soft and shower partons in the 
intermediate $p_{\mathrm T}$ region. Also, a phenomenological 
approach, called EPOS~\cite{epos}, based on a parton model, has 
described the $d$+Au collision data at RHIC. In this model 
the nuclear effects are included through elastic and inelastic parton 
ladder splitting. 

It is of interest to see how these models compare to
data for rapidity asymmetry of identified hadrons from $d$+Au collisions. 
More precisely, identified hadron $Y_{\mathrm{Asym}}$, a 
more differential quantity, 
may allow some determination of the relative
contribution of the  physical processes discussed above.
Strong particle type (baryon and meson) dependence of the 
nuclear modification factor and azimuthal anisotropy at 
intermediate $p_{\mathrm T}$ (2~$<$~$p_{\mathrm T}$~$<$~6 GeV/$c$) 
has been observed in Au+Au collisions at RHIC~\cite{baryon_meson}. 
The present study will investigate if such particle type 
(baryon and meson) dependence is observed in $Y_{\mathrm{Asym}}$
for $d$+Au collisions.

In addition to providing insight into different particle production 
mechanisms at forward and backward rapidity for $d$+Au collisions, 
the measurements presented here may be used to study the presence of possible 
effects of valence quarks and isospin on particle production. 
At high $p_{\mathrm T}$ and rapidities away from midrapidity,
the role of valence quarks becomes increasingly dominant. 
Such studies are even more interesting for $n$-tag events 
(events where the neutron in the deuteron does not interact with 
the gold nucleus). 
A comparative study between $p$+Au 
($n$-tag) and $d$+Au data is of interest. For $n$-tag events at forward rapidity 
and high $p_{\mathrm T}$, the two valence $u$ quarks 
in the proton of the deuteron should lead to more production 
of $\pi^{+}$ ($u\bar{d}$) compared to $\pi^{-}$ ($d\bar{u}$). 
For backward rapidities, if the flavor 
distribution in sea quarks is uniform and the incoming gold nucleus 
has no asymmetry in $u$ and $d$ quarks, one expects the ratio 
$\pi^{-}/\pi^{+}$ $\sim$ 1. This difference between forward and 
backward rapidity may be more pronounced for $\bar{p}/p$. 
Study of particle ratios as a function of rapidity at high $p_{\mathrm T}$ 
in peripheral and $n$-tag events for $d$+Au collisions may provide some 
information on the flavor dependence of particle production. These ratios are
in principle sensitive to the fragmentation function ratios of $u$-quarks
to $\pi^{-}$ and $\pi^{+}$~\cite{emc}, to the ratio of 
$(u,d)$-quarks fragmenting to protons~\cite{opal}, and to the fractional
contributions of quarks and gluons to hadrons at the given momentum.

In this paper, we present the first results for 
the rapidity asymmetry of charged pion, proton
and anti-proton production at high $p_{\mathrm T}$ for $d$+Au
collisions at $\sqrt{s_{\mathrm {NN}}}$ = 200 GeV measured
by the STAR experiment~\cite{starnim} at RHIC. A similar study
for inclusive charged hadrons has been reported in Ref.~\cite{star_asym}.
The asymmetry is studied as a function of $p_{\mathrm T}$ for 
different collision centralities in the two rapidity regions 
$\mid$$y$$\mid$~$<$~0.5 and  0.5~$<$~$\mid$$y$$\mid$~$<$~1.0. 
In section II we discuss the detectors used in the analysis, trigger 
and centrality selection, particle identification
at high $p_{\mathrm T}$, 
and the systematic errors. In section III we 
discuss the rapidity, $p_{\mathrm T}$, 
species, and centrality dependence of $Y_{\mathrm{Asym}}$. In section IV,
the $Y_{\mathrm{Asym}}$ results are compared to 
calculations from various models discussed earlier. 
In section V, we present the rapidity dependence of the nuclear modification 
factor for $\pi^{+}+\pi^{-}$ and $p$+$\bar{p}$ . In section VI, we study 
the anti-particle to particle ratios as a function of rapidity at 
high $p_{\mathrm T}$ in $n$-tag and 
peripheral $d$+Au events in order to investigate the 
flavor dependence of particle production.
Section VII completes this work with the summary of our findings.

\section{EXPERIMENT AND ANALYSIS}

\subsection{Detectors}

For the present analysis we use data recorded by the Time Projection 
Chamber (TPC)~\cite{startpc} in the STAR experiment at RHIC. 
The TPC is STAR's primary tracking device. 
It is 4.2 m long  and 4 m in diameter. The sensitive volume of the TPC  
contains P10 gas (10\% methane, 90\% argon) regulated at
2 mbar  above atmospheric pressure. 
The TPC data are used to determine particle trajectories, momenta, and particle-type
through ionization energy loss
($dE/dx$). Its  acceptance 
covers $\pm1.8$ units of pseudorapidity ($\eta$) and  the full azimuthal angle. 
Charged particle momenta are determined from the TPC data for the $d$+Au
run in the year 2003 in which STAR's solenoidal magnet field was set to 0.5\,T. 
Two Zero Degree Calorimeters (ZDCs)~\cite{starzdc} 
situated along both sides of the beam axis, about 18~m from the nominal 
collision point (center of TPC), were used for triggering. 
The collision centrality is obtained from 
the charged hadron multiplicity measured by STAR's Forward Time Projection
Chambers (FTPCs)~\cite{starftpc}. The details of the design and other 
characteristics of the detectors can be found in Ref.~\cite{starnim}. 
The details of the trigger condition, collision centrality selection, and 
method of high $p_{\mathrm T}$ particle identification  are described 
below.

\subsection{Trigger conditions}
The ZDC in the Au beam direction, which is assigned negative 
pseudorapidity ($\eta$), was used as the trigger detector for 
obtaining the minimum bias data. The minimum bias trigger required 
at least one beam-rapidity neutron in the ZDC. The trigger efficiency 
was found to be 95$\pm$3\% of the $d$+Au hadronic cross section 
$\sigma_{hadr}^{dAu}$. Trigger backgrounds were determined 
using data recorded for beam crossings without collisions. For the $n$-tag events, 
the ZDC in the deuteron beam direction was used. Such events 
were required to have at least one beam rapidity neutron 
in the ZDC. The cross section for such a process  
was measured to be (19.2$\pm$1.3)\% 
of $\sigma_{hadr}^{dAu}$. The vertex was reconstructed for 
93$\pm$1\% of triggered minimum bias events. 
A total of 11.7 million minimum bias $d$+Au events 
and 2.0 million $n$-tag events having a vertex within
$\pm$ 30 cm of the nominal interaction point along the beam direction were analyzed. 
Two rapidity regions were used: $\mid$$y$$\mid$~$<$~0.5 
and 0.5~$<$~$\mid$$y$$\mid$~$<$~1.0, and  the $p_{\mathrm T}$ range 
was 2.5~$<$~$p_{\mathrm T}$~$<$~10~GeV/$c$.
The $p_{\mathrm T}$ spectra were corrected for trigger 
and vertex-finding inefficiencies. Further details of trigger 
conditions for the minimum bias data can be found in 
Ref.~\cite{star_rdau}. 

\begin{table}
\caption{ Centrality selection, number of participating nucleons, and
number of binary collisions for $d$+Au collisions at 
$\sqrt{s_{\mathrm {NN}}}$ = 200 GeV.
\label{table1}}
\begin{tabular}{cccccc}
\tableline
\% cross section&$N_{\mathrm{chtrk}}^{\mathrm{FTPC}}$ &$\langle N_{\mathrm{part}} \rangle$
&$\langle N_{\mathrm{bin}} \rangle$\\
\tableline
0--20   &  $>$ 17   & 15.67 $\pm$ 1.07 & 15.1 $\pm$ 1.15\\
20--40  &  11--17   & 11.16 $\pm$ 1.25 & 10.6 $\pm$ 1.38\\
40--100 &  $<$ 11   & 5.14  $\pm$ 0.47 & 4.2  $\pm$ 0.51\\
0--100  &  $>$ 0    & 8.31  $\pm$ 0.34 & 7.5  $\pm$ 0.38\\

\tableline
\end{tabular}
\end{table}
\subsection{Collision centrality }
Uncorrected charged track multiplicity 
($N_{\mathrm{chtrk}}^{\mathrm{FTPC}}$)
measured within 
-3.8~$<$~$\eta$~$<$~-2.8 by the FTPC 
was used to determine the collision centrality for $d$+Au collisions.
The centrality selection criteria is given in Table~\ref{table1}, 
along with the average number of binary collisions ($N_{\mathrm bin}$) 
and the number of participating nucleons ($N_{\mathrm part}$) 
estimated using a Monte Carlo Glauber calculation~\cite{glauber} 
incorporating the Hulth\'{e}n wave function of 
the deuteron~\cite{hulthen}. In this model 
$\sigma_{hadr}^{dAu}$=2.21$\pm$0.09 b, and $N_{\mathrm bin}$ 
for $n$-tag events is 2.9$\pm0.2$. This model provides
reasonable agreement with the measured charged track multiplicity
distribution in the FTPC and the single neutron cross section
measured by the ZDC on the deuteron side.
Further details of centrality tagging in $d$+Au collisions
can be found in  Ref.~\cite{star_rdau}.

\subsection{Particle identification at high $p_{\mathrm T}$ }
\begin{figure}
\begin{center}
\includegraphics[scale=0.45]{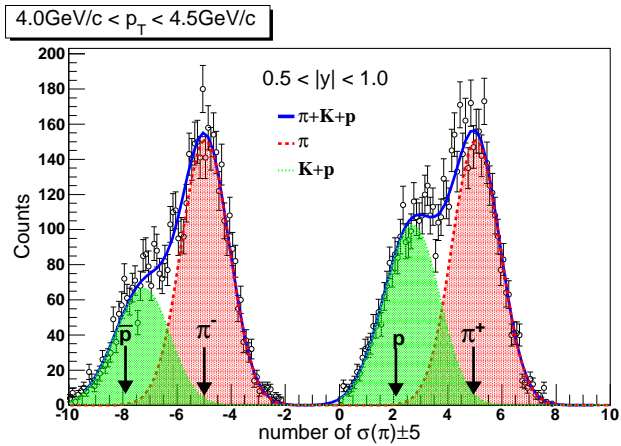}
\caption{(Color online) $dE/dx$ distribution normalized by pion 
$dE/dx$ at 4.0~$<$~$p_{\mathrm T}$~$<$~4.5~GeV/$c$ and 
0.5~$<$~$\mid$$\eta$$\mid$~$<$~1.0, shifted by $\pm$5 
for positive and negative charged particles, respectively. 
The distributions are for minimum bias $d$+Au collisions. 
The pion, proton, and anti-proton peak positions are 
indicated by arrows.
}
\label{fig0}
\end{center}
\end{figure}

Particle identification at high transverse momenta 
($p_{\mathrm T}$~$>$~2.5 GeV/$c$) is done by exploiting 
the relativistic rise of the ionization energy loss. 
Here we briefly describe the identification procedure (see Ref.~\cite{rdEdx,pid}).
For $2.5~ < ~p_T ~{}^{<}_{\sim}~ 10$~ GeV/$c$, there is a difference of
about 10--20\% between the pion $dE/dx$ and the $dE/dx$ for kaons and protons,
due to the relativistic rise of the ionization energy loss for pions. 
This results in a few sigma (1-3$\sigma$) separation. 
The $dE/dx$ resolution is $\sim8\%$~\cite{rdEdx}.

Pions are the dominant component of the hadron yield for $d$+Au collisions at RHIC. 
The prominent peak in the $dE/dx$ distribution is used to determine the 
pion yield in this $p_{\mathrm T}$ range.
To extract the pion yield in a given $p_{\mathrm T}$ bin, we 
performed a six Gaussian fit to the normalized $dE/dx$ distributions of
positive and negative hadrons simultaneously.
The normalized $dE/dx$ in general is defined as 
$n\sigma_{X}^{Y}=log((dE/dx)_Y/B_{X})/\sigma_{X}$, where $X,Y$
can be $e^{\pm},\pi^{\pm},K^{\pm}$ or $p(\bar{p})$. $B_{X}$ is the
expected mean $dE/dx$ of  particle $X$, and $\sigma_{X}$ is the
$dE/dx$ resolution of the TPC. 
Fig.~\ref{fig0} shows a typical $dE/dx$ distribution normalized to pion $dE/dx$
(referred to as the $n\sigma_{\pi}$ distribution) for
charged hadrons with 4.0~$<$~$p_{\mathrm T}$~$<$~4.5~GeV/$c$ and 
0.5 $<$ $\mid$$\eta$$\mid$ $<$ 1.0. 
For clarity of presentation, the $n\sigma_{\pi}$ distributions in 
Fig.~\ref{fig0} are shifted by $\pm$5 for positive and negative charged 
particles, respectively. The $n\sigma_{\pi}^{\pi}$ distribution is a 
normal Gaussian distribution with an ideal calibration. 
The six Gaussians are for $\pi^{\pm}$,
$K^{\pm}$ and $p(\bar{p})$. The $n\sigma_{\pi}^{K}-n\sigma_{\pi}^{\pi}$
and $n\sigma_{\pi}^{p(\bar{p})}-n\sigma_{\pi}^{\pi}$ values
are estimated by studying the $n\sigma_{\pi}^{\pi}-n\sigma_{K}^{K}$,
$n\sigma_{\pi}^{\pi}-n\sigma^{p}_{p}$, and $n\sigma_{\pi}^{h^{+}}$- 
$n\sigma_{\pi}^{h^{-}}$
distributions. The widths of the six Gaussians are taken to 
be the same. 
The Gaussian distribution used to extract the 
pion yield and the pion, proton and 
anti-proton peak positions are also shown in the figure.

\begin{figure*}
\begin{center}
\includegraphics[scale=0.75]{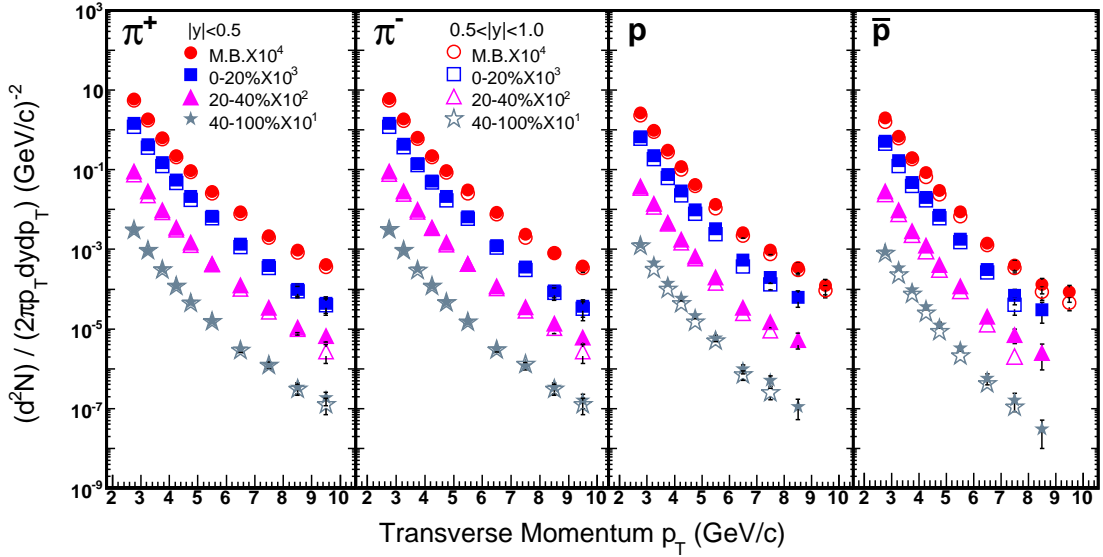}
\caption{(Color online) High transverse momentum spectra 
($p_{\mathrm T}$~$>$~2.5 GeV/$c$) of charged pions, proton, and 
anti-proton for the rapidity regions $\mid$$y$$\mid$~$<$~0.5 
(solid symbols) and 0.5~$<$ $\mid$$y$$\mid$~$<$~1.0 (open symbols) 
for $d$+Au collisions and various event centrality classes 
at $\sqrt{s_{\mathrm {NN}}}$ = 200 GeV. 
}
\label{fig1}
\end{center}
\end{figure*}
The proton yield is obtained by integrating
the entries ($Y$) in the low part of the $dE/dx$ 
distribution, about $2.5\sigma$ away from the pion 
$dE/dx$ peak. The integration limits were varied 
to check the stability of the results. 
The kaon contamination is estimated via either of 
the equations given below. The raw  proton yield is
$$p = (Y-\beta (h-\pi))/{(\alpha-\beta)}$$ or,
$$p = (Y-\beta K^{\mathrm 0}_{\mathrm S})/{\alpha},$$ where $\alpha$
and $\beta$ are the proton and kaon efficiencies from the integration
described above, derived from the $dE/dx$ calibration, resolution, and
the Bichsel function~\cite{rdEdx,bic}.  In the first case the kaon 
contamination is estimated through the yields of the inclusive 
hadrons ($h$) and pions; in the second case using
$K^{\mathrm 0}_{\mathrm S}$ measurements~\cite{pid}
(only available for $\mid$$y$$\mid$~$<$~0.5 up to $p_{\mathrm T}$~$<$~5 GeV/c).  
The typical values of $\alpha$ for a $dE/dx$ cut slightly away 
from the proton peak position is 0.4. The $\beta$ values 
decrease from 0.2 to 0.08 with $p_{\mathrm T}$ in the range 
2.5~$<$~$p_{\mathrm T}$~$<$~10 GeV/$c$. 
At high $p_{\mathrm T}$, the yields of other stable particles 
(i.e., electrons and deuterons) are at least two orders of magnitude 
smaller than those of pions and are negligible for our studies.  
The two results are consistent in the region where STAR $K^{\mathrm 0}_{\mathrm S}$ 
measurements are available. 
Since the energy loss of particles in the TPC is almost independent 
of charge sign, the dependence of 
$h^{-}/h^{+}$ on $n\sigma_{\pi}$ is due to different particle 
composition and the $dE/dx$ separation between pion,
kaon and proton~\cite{rdEdx}. This provided a
consistency check for the yields.

The $dE/dx$ resolution is better for longer tracks, 
shorter drift distance, stronger magnetic field, 
smaller multiplicity and lower beam luminosity.
Due to longer tracks and shorter drift
distances for particles produced at higher $y$, the $dE/dx$
resolution gets better. Thus, the separations between pions and
kaons or (anti-)protons were larger 
for 0.5~$<$~$\mid$$y$$\mid$~$<$~1.0 than for 
$\mid$$y$$\mid$~$<$~0.5~\cite{pid}, and particle
identification is easier at larger $p_{\mathrm T}$.

\begin{table}
\caption{ Correction factors for identified hadron spectra at 
high $p_{\mathrm T}$ ($>$~2.5 GeV/$c$) for minimum bias $d$+Au collisions.
\label{table2}}
\begin{tabular}{ccc}
\tableline
Type&\%  \\
\tableline
Trigger efficiency              &  95 $\pm$ 3      \\
Vertex  efficiency              &  93 $\pm$ 1       \\
Track reconstruction efficiency &  $\sim$ 90 $\pm$ 8 \\
($\mid$$y$$\mid$ $<$ 0.5)         &             \\
Track reconstruction efficiency &  $\sim$ 82 $\pm$ 8  \\
(0.5 $<$ $\mid$$y$$\mid$ $<$ 1.0) &              \\
Background contamination        &  $\sim$ 5  $\pm$ 1  \\
\tableline
\end{tabular}
\end{table}

\begin{table*}
\caption{Systematic errors for identified hadron minimum bias yields at 
high $p_{\mathrm T}$ ($>$~2.5 GeV/$c$) for $d$+Au collisions. 
\label{table3}}
\begin{tabular}{ccc}
\tableline
Sources of uncertainty                 & \% Error \\
\tableline
Modeling detector response             &  8       \\
Momentum resolution                    &  4 (at $p_{\mathrm T}$ = 7 GeV/c) \\
Spatial distortion                     &  8  \\
$dE/dx$ pion peak position               &  8   \\
$dE/dx$ proton peak position             &  8    \\
Kaon contamination to proton yield     &  12 (at $p_{\mathrm T}$ = 7 GeV/c)\\
Protons from hyperon decay             &  7   (at $p_{\mathrm T}$ = 7 GeV/c)\\
Normalization (trigger and luminosity) & 10 \\
\tableline
\end{tabular}
\end{table*}
\subsection{Correction factors }
The various correction factors for the identified hadron spectra
are listed in Table~\ref{table2}. The trigger and vertex
efficiencies were discussed previously. The identified hadron
track reconstruction efficiency was estimated by embedding Monte Carlo
particles into the real data and then following the full reconstruction 
procedure. It was observed to be independent of $p_{\mathrm T}$ for 
$p_{\mathrm T}$~$>$~2.5 GeV/c for both  rapidity regions.
The reconstruction efficiency for $p_{\mathrm T}$~$>$~2.5~GeV/$c$ 
for charged pions and protons are $\sim$ 92\% and $\sim$ 90\%,
respectively, in the rapidity region $\mid$$y$$\mid$ $<$ 0.5. 
For 0.5~$<$~$\mid$$y$$\mid$~$<$~1.0, the reconstruction efficiency
for charged pions and protons is $\sim$ 82\% and 84\%, respectively. 
The background contamination in the pion spectra for 
$p_{\mathrm T}$~$>$~2.5 GeV/$c$, primarily from 
$K_{\mathrm 0}^{\mathrm S}$ weak decay,
is $\sim$ 5\%. No strong centrality dependence was observed 
in the correction factors. 
The charged pion, proton and anti-proton spectra are corrected 
for efficiency and background effects. The inclusive proton and 
anti-proton spectra are presented without 
hyperon feed down corrections~\cite{pid,star_tof}. 
Preliminary study shows that the ratio of $\Lambda$ to inclusive $p$ in 
the rapidity range $\mid$$y$$\mid$~$<$~0.5 decreases from 0.7 to 0.3 
with increase in $p_{\mathrm T}$ from 2.5 GeV/$c$ to 5.5 GeV/$c$.

\subsection{Systematic errors }
The total systematic uncertainties associated with the pion yields 
are estimated to be ${}^<_\sim$ 15\%, and those for proton and 
anti-proton yields are estimated to be ${}^<_\sim$ 22\%. 
They are of similar order for both the rapidity regions, and the
average values for minimum bias collisions are given in 
Table~\ref{table3}.

The sources of systematic error on the high  $p_{\mathrm T}$ 
yield arise owing to: (a) uncertainty in modeling
the detector response in the Monte Carlo simulations, (b)
momentum resolution (increases with $p_{\mathrm T}$)~\cite{pid}, (c) 
difference in the yields for different
TPC sectors as a result of spatial distortion effects, (d) uncertainty 
in determining the pion and proton $dE/dx$ peak positions,
(e) uncertainty
in estimating the kaon contamination to proton yields
(increases from 7\% at $p_{\mathrm T}$~=~2.5~GeV/$c$ to 15\% at
$p_{\mathrm T}$ = 10 GeV/$c$), and (f) uncertainty due to
protons from hyperon decay that are reconstructed as primordial protons
at a slightly higher $p_{\mathrm T}$ than their true value, with
a worse momentum resolution (increases from 2\% at 
$p_{\mathrm T}$~=~2.5 GeV/$c$ to 10\% at $p_{\mathrm T}$ = 10 GeV/$c$).
There is an additional 10\%~\cite{star_rdau} normalization uncertainty due to
trigger and luminosity uncertainties.
These systematic errors are not shown in Figure~\ref{fig1}.

As this work focuses mainly on ratios such as $Y_{\mathrm{Asym}}$, 
most of the systematic errors cancel. The resultant systematic
error on $Y_{\mathrm{Asym}}$ is about 5\%. The errors shown for  
figures with ratios are statistical and systematic errors added in quadrature.

Figure~\ref{fig1} shows the measured invariant yields of 
charged pions, protons, and anti-protons for the 
$p_{\mathrm T}$ range 2.5~$<$~$p_{\mathrm T}$~$<$~10 GeV/$c$ 
in the rapidity regions, $\mid$$y$$\mid$~$<$~0.5 (solid 
symbols) and 0.5~$<$~$\mid$$y$$\mid$~$<$~1.0 (open symbols)
for minimum bias and various collision centrality classes 
for $d$+Au collisions at $\sqrt{s_{\mathrm {NN}}}$~=~200~GeV. 
The $p_{\mathrm T}$ spectra are corrected for the trigger, vertex
and reconstruction efficiencies and the background effects listed 
in Table~\ref{table2}.

\section{RAPIDITY ASYMMETRY}

In this section we discuss the $y$, $p_{\mathrm T}$, species
and centrality dependence of $Y_{\mathrm {Asym}}$.

\subsection{Rapidity, transverse momentum, and species dependence}
\begin{figure}
\begin{center}
\includegraphics[scale=0.4]{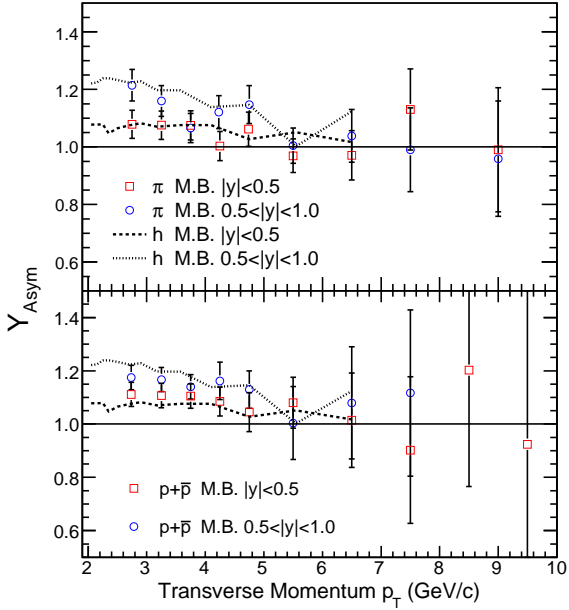}
\caption{(Color online)
High transverse momentum rapidity asymmetry factor ($Y_{\mathrm {Asym}}$) for 
$\pi^{+}$+$\pi^{-}$ and $p$+$\bar{p}$ for $\mid$$y$$\mid$~$<$~0.5
and 0.5~$<$~$\mid$$y$$\mid$~$<$~1.0 for minimum bias $d$+Au collisions at
$\sqrt{s_{\mathrm {NN}}}$ = 200 GeV. For comparison the inclusive
charged hadron results from STAR~\cite{star_asym} are also shown by the curves.
}
\label{fig2}
\end{center}
\end{figure}

Figure~\ref{fig2} shows the high $p_{\mathrm T}$ dependence of 
$Y_{\rm{Asym}}$ for $\pi^{+}$+$\pi^{-}$ and $p$+$\bar{p}$ 
for the rapidity regions $\mid$$y$$\mid$~$<$~0.5  and 
0.5~$<$~$\mid$$y$$\mid$~$<$~1.0  
for minimum bias events.
The backward rapidity is considered as the gold-side 
and corresponds to the negative rapidity region. 
The forward rapidity is the deuteron-side and
corresponds to positive rapidity. 
For comparison the pseudorapidity asymmetry for charged 
hadrons~\cite{star_asym} is
shown also in the figure. The following observations are made: \\
(a) $Y_{\rm{Asym}}$ is found to be larger for  
0.5~$<$~$\mid$$y$$\mid$~$<$~1.0 than for $\mid$$y$$\mid$ $<$ 0.5
for all the hadrons with 2.5~$<$~$p_{\mathrm T}$~$<$~5 GeV/$c$. 
This may indicate the presence of some rapidity dependence of 
nuclear effects such as parton saturation, nuclear shadowing, 
or energy loss in cold nuclear matter. \\
(b) The $Y_{\rm{Asym}}$ values are consistent with unity for both rapidity 
regions at high $p_{\mathrm T}$ ($>$~5.5 GeV/$c$), suggesting
absence of nuclear effects on particle production in $d$+Au collisions
for this $p_{\mathrm T}$ range. \\
(c) $Y_{\rm{Asym}}$ for charged pions is greater than unity and 
decreases monotonically with increasing 
$p_{\mathrm T}$ for 2.5~$<$~$p_{\mathrm T}$~$<$~5 GeV/$c$. 
Although $Y_{\rm{Asym}}$ for $p$+$\bar{p}$ is also greater than unity,
the trend seems to be towards a constant value in this $p_{\mathrm T}$ range. 
These features are opposite to predictions from models based on 
incoherent initial multiple partonic scattering and independent
fragmentation~\cite{wang}. Such models predict that $Y_{\rm{Asym}}$ 
is less than  unity at intermediate  $p_{\mathrm T}$ and approximately unity 
for larger $p_{\mathrm T}$~\cite{star_asym}. \\
(d) For $\mid$$y$$\mid$~$<$~0.5, $Y_{\rm{Asym}}$ for $p$+$\bar{p}$ is 
slightly larger than it is for charged pions for 2.5~$<$~$p_{\mathrm T}$~$<$~4 GeV/$c$. 
For 0.5 $<$ $\mid$$y$$\mid$ $<$ 1.0  no strong particle type 
dependence is observed for $Y_{\rm{Asym}}$. This is in contrast to the observed
baryon-meson differences for the same $p_{\mathrm T}$ range for Au+Au
collisions, which were described by recombination-based models~\cite{reco}.

\subsection{Centrality dependence}
\begin{figure}
\begin{center}
\includegraphics[scale=0.4]{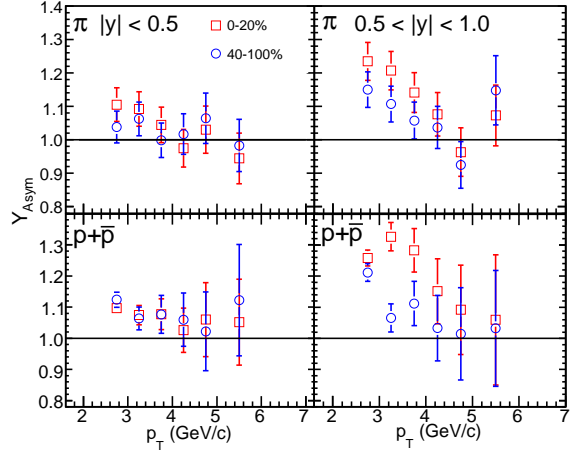}
\caption{(Color online)
Centrality dependence of high transverse momentum rapidity 
asymmetry factor ($Y_{\mathrm {Asym}}$) for 
$\pi^{+}$+$\pi^{-}$ and $p$+$\bar{p}$ at $\mid$$y$$\mid$~$<$~0.5 (left panels)
and 0.5~$<$~$\mid$$y$$\mid$~$<$~1.0 (right panels) for $d$+Au collisions at 
$\sqrt{s_{\mathrm {NN}}}$ = 200 GeV. 
}
\label{fig3}
\end{center}
\end{figure}
In Fig.~\ref{fig3} we show the centrality dependence 
of $Y_{\mathrm {Asym}}$ at high $p_{\mathrm T}$ 
for $\pi^{+}$+$\pi^{-}$ and $p$+$\bar{p}$ 
for the two rapidity regions $\mid$$y$$\mid$~$<$~0.5 (left panels) and 
0.5~$<$~$\mid$$y$$\mid$~$<$~1.0 (right panels).
The data are shown only for 
2.5~$<$ $p_{\mathrm T}$~$<$~5.5 GeV/$c$. The $Y_{\mathrm {Asym}}$ 
values approach unity for the centrality classes studied 
for $p_{\mathrm T}$~$>$~5.5 GeV/$c$ in both the rapidity regions.
For $\mid$$y$$\mid$~$<$~0.5, a prominent centrality dependence of 
$Y_{\mathrm {Asym}}$ is not observed. For
0.5~$<$~$\mid$$y$$\mid$~$<$~1.0, $Y_{\mathrm {Asym}}$
is larger for central (0--20\%) compared to peripheral (40--100\%) 
events for 2.5~$<$~$p_{\mathrm T}$~$<$~4 GeV/$c$. 
The indication of a centrality dependence in the $Y_{\mathrm {Asym}}$
at 0.5~$<$~$\mid$$y$$\mid$~$<$~1.0 is consistent with predictions 
from saturation models~\cite{cgc}. However in such models the centrality 
dependence is much stronger than 
observed in the present data~\cite{star_asym}.

\section{MODEL COMPARISON}
In this section we compare the measured high $p_{\mathrm T}$ 
identified hadron $Y_{\mathrm {Asym}}$ in the rapidity regions 
$\mid$$y$$\mid$~$<$~0.5 and 0.5~$<$~$\mid$$y$$\mid$~$<$~1.0 for 
minimum bias $d$+Au collisions at $\sqrt{s_{\mathrm {NN}}}$~=~200 GeV 
with predictions from various models (Figs.~\ref{fig4}--\ref{fig8}).

\subsection{Comparison to the nuclear shadowing model}

First we compare the high $p_{\mathrm T}$ charged 
pion $Y_{\mathrm {Asym}}$  
in both  rapidity regions with model predictions that
incorporate only nuclear shadowing~\cite{vogt}. 
In these calculations two  parameterizations of nuclear 
shadowing, covering the extremes of gluon shadowing at low $x$, 
are taken. 
The parametrization by Eskola {\it et al.}~\cite{eks} is 
referred to as EKS98. The other, FGS, is from Frankfurt, Guzey, and Strikman~\cite{fgs} 
(FGSO, the original parametrization, along with FGSH and FGSL 
for high and low gluon shadowing). The calculations use MRST 
leading order (LO) parton distribution functions~\cite{mrst}. The 
fragmentation of  produced partons into charged pions uses the 
LO Kniehl-Kramer-Potter (KKP) fragmentation functions~\cite{kkp} 
obtained from a fit to $e^{+}$+$e^{-}$ data. In EKS98 the valence 
quark shadowing is identical for $u$ and $d$ quarks at the minimum 
momentum scale of the hard interaction. In FGS the  
EKS98 valence quark shadowing ratios are used as input,
along with Gribov theory and hard diffraction.
The charged hadron $R_{\mathrm {dAu}}$ was reasonably well-described 
by such a model using the FGS parametrization~\cite{vogt}.

\begin{figure}
\begin{center}
\includegraphics[scale=0.42]{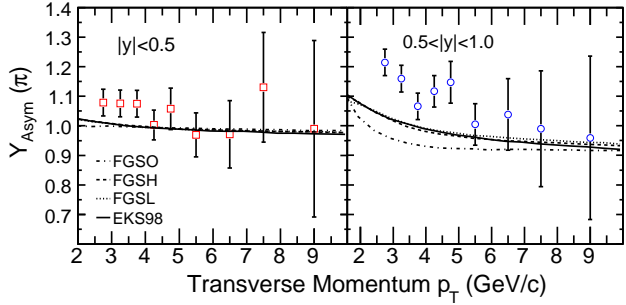}
\caption{(Color online)
High transverse momentum rapidity 
asymmetry factor ($Y_{\mathrm {Asym}}$) for 
$\pi^{+}$+$\pi^{-}$ at $\mid$$y$$\mid$ $<$ 0.5 
and 0.5 $<$ $\mid$$y$$\mid$ $<$ 1.0 for minimum bias $d$+Au collisions at 
$\sqrt{s_{\mathrm {NN}}}$ = 200 GeV compared to a model with only
nuclear shadowing~\cite{vogt}. The different curves represent
different parametrization of nuclear shadowing. See text for more details.
}
\label{fig4}
\end{center}
\end{figure}

Our charged pion data (Fig.~\ref{fig4}) indicate that nuclear shadowing as
implemented in the models discussed, cannot explain the
measured $Y_{\mathrm {Asym}}$ for 2.5~$<$~$p_{\mathrm T}$~$<$~5 GeV/$c$
for both $\mid$$y$$\mid$~$<$~0.5 and 0.5~$<$~$\mid$$y$$\mid$~$<$~1.0.
The differences between data and model increase with  
increasing rapidity. 
At larger $p_{\mathrm T}$, the data values approach unity, 
indicating an absence of nuclear effects. The effect 
on $Y_{\mathrm {Asym}}$ of using 
a different parametrization of 
nuclear shadowing at high 
$p_{\mathrm T}$ is found to be negligible for $\mid$$y$$\mid$~$<$~0.5. 
However, some differences are observed 
in FGS for 0.5~$<$~$\mid$$y$$\mid$~$<$~1.0.

$Y_{\mathrm {Asym}}$ from the nuclear shadowing model covering 
the extremes of gluon shadowing at low $x$ is not consistent
with the measured values. The comparison therefore 
provides an idea of the maximum contribution to $Y_{\mathrm {Asym}}$
from only nuclear shadowing in $d$+Au collisions at 
$\sqrt{s_{\mathrm {NN}}}$ = 200.

\subsection{Comparison to the multiple scattering+shadowing+energy loss model}
\begin{figure}
\begin{center}
\includegraphics[scale=0.42]{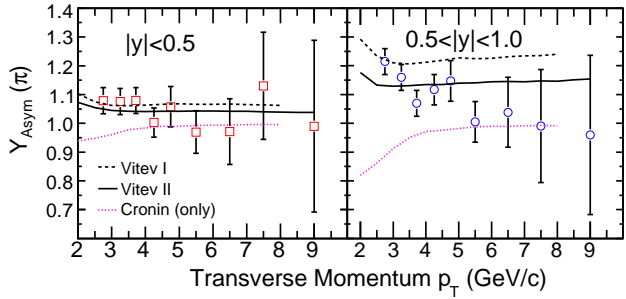}
\caption{(Color online)
High transverse momentum rapidity 
asymmetry factor ($Y_{\mathrm {Asym}}$) for 
$\pi^{+}$+$\pi^{-}$ at $\mid$$y$$\mid$ $<$ 0.5 
and 0.5 $<$ $\mid$$y$$\mid$ $<$ 1.0 for minimum bias $d$+Au collisions at 
$\sqrt{s_{\mathrm {NN}}}$ = 200 GeV compared to models
incorporating multiple scattering, shadowing, and energy
loss in cold nuclear matter~\cite{vitev}. See text for more details.
}
\label{fig5}
\end{center}
\end{figure}

Next we compare the high $p_{\mathrm T}$ charged 
pion $Y_{\mathrm {Asym}}$ to a model that includes only 
coherent multiple scattering, which leads to transverse 
momentum broadening (Cronin effect), and to calculations with 
the addition of power corrections (dynamical shadowing) and energy 
loss in cold nuclear matter~\cite{vitev}. In this model a systematic 
calculation of the coherent multiple parton scattering in 
$p$+A collisions is carried out in terms of the perturbative QCD 
factorization approach. It also incorporates initial state 
parton energy loss in the perturbative calculations. 
We observe (Fig.~\ref{fig5}) that in both rapidity regions 
model expectations from the Cronin effect, 
for 2.5~$<$~$p_{\mathrm T}$~$<$~5~GeV/$c$, are in qualitative disagreement 
with the data. This indicates that multiple scattering is not the source of the 
observed asymmetry. 
Fig.~\ref{fig5} shows also a comparison of the charged pion data with results of 
calculations which incorporate multiple scattering, 
dynamical shadowing, and 
a varying degree of energy loss in cold nuclear matter. 
The calculation, labeled as Vitev-I, has a slightly larger effective 
energy loss in cold nuclear matter than the one labeled  Vitev-II.
For $\mid$$y$$\mid$~$<$~0.5, both the Vitev-I and Vitev-II results 
are in reasonable agreement with the data within errors.
For 0.5~$<$~$\mid$$y$$\mid$~$<$~1.0, the Vitev-I result 
slightly overpredicts the measured $Y_{\mathrm {Asym}}$. The model 
calculations beyond $p_{\mathrm T}$~$>$~3 GeV/$c$ are independent 
of $p_{\mathrm T}$, while the measured $Y_{\mathrm {Asym}}$ tends 
to decrease with $p_{\mathrm T}$.

For the Vitev models the rapidity dependence of 
$Y_{\mathrm {Asym}}$ seems to be sensitive to effective 
energy loss. The decrease in $Y_{\mathrm {Asym}}$ with 
 $p_{\mathrm T}$ in the model is restricted to 
$p_{\mathrm T}~<~2.5$~GeV/$c$. 
It will be important to have predictions from this model 
for proton and anti-protons to investigate the 
possible particle species dependence of multiple scattering 
in $d$+Au collisions.

\subsection{Comparison to the EPOS model}
\begin{figure}
\begin{center}
\includegraphics[scale=0.45]{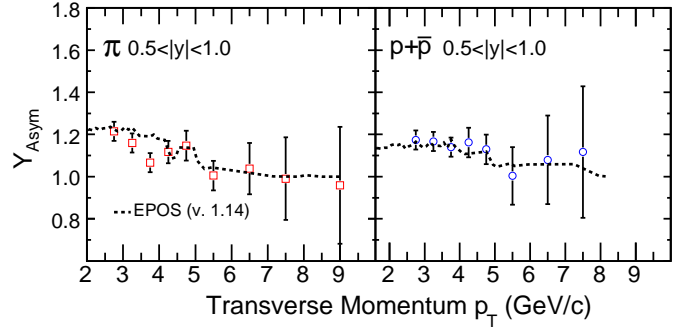}
\caption{(Color online)
High transverse momentum rapidity 
asymmetry factor ($Y_{\mathrm {Asym}}$) for 
$\pi^{+}$+$\pi^{-}$ and  $p$+$\bar{p}$ for 
0.5~$<$~$\mid$$y$$\mid$~$<$~1.0 for minimum bias $d$+Au collisions at 
$\sqrt{s_{\mathrm {NN}}}$ = 200 GeV compared to the EPOS model~\cite{epos}.
See text for more details.
}
\label{fig6}
\end{center}
\end{figure}

In Fig.~\ref{fig6} the measured  $Y_{\mathrm {Asym}}$ for 
$\pi^{+}$+$\pi^{-}$ and $p$+$\bar{p}$ for 0.5 $<$ $\mid$$y$$\mid$ $<$ 1.0 
are compared to the results from the EPOS model.
In the EPOS model~\cite{epos} elastic and inelastic parton ladder 
splitting are the key processes. A parton ladder refers to the 
dynamical process of parton--parton scattering with successive 
emission of partons. The emission process can be an initial state, 
space-like cascade, or final state, time-like cascade. The elastic 
splitting in this model can be related to screening and saturation,
while inelastic splitting is related to the hadronization process. 
This phenomenological model has been very successful in describing 
the inclusive charged hadron $d$+Au data~\cite{epos}. 

The EPOS model predictions (v. 1.14) are consistent with
the measured $Y_{\mathrm {Asym}}$ values for both charged 
pions and $p$+$\bar{p}$.

\subsection{Comparison to the recombination model}
\begin{figure}
\begin{center}
\includegraphics[scale=0.45]{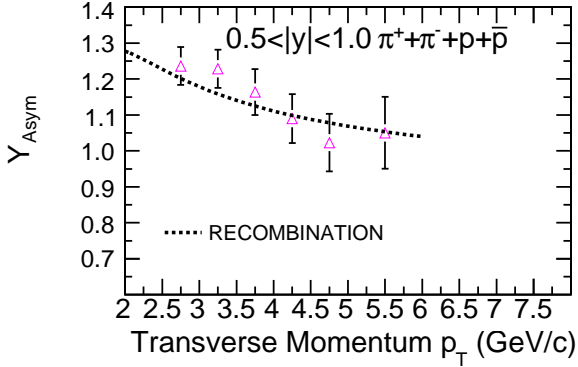}
\caption{(Color online)
High transverse momentum rapidity 
asymmetry factor ($Y_{\mathrm {Asym}}$) for 
$\pi^{+}$+$\pi^{-}$+$p$+$\bar{p}$ and 
0.5 $<$ $\mid$$y$$\mid$ $<$ 1.0 in 0--20\% central $d$+Au collisions at 
$\sqrt{s_{\mathrm {NN}}}$ = 200 GeV compared to the 
recombination model~\cite{hwa}.
See text for more details.
}
\label{fig7}
\end{center}
\end{figure}

The recombination model reproduces 
some of the observed features of RHIC data~\cite{hwa,reco}.
It successfully describes the Cronin effect for $d$+Au data
without any need for $k_{\mathrm T}$ broadening in initial state 
interactions. 
Although questions have been raised concerning issues such as 
decrease in entropy of the system and the spatial extent of 
the recombining subsystems, it is useful to
compare the experimental measurements with this model to investigate
the relative importance of various physical processes. 

Fig.~\ref{fig7} compares model predictions with the
measured $Y_{\mathrm {Asym}}$  for 
$\pi^{+}$+$\pi^{-}$+$p$+$\bar{p}$ for the 
rapidity region 0.5 $<$ $\mid$$y$$\mid$ $<$ 1.0 in 0--20\% central $d$+Au collisions. 
The model predictions from Ref.~\cite{hwa}
are consistent with the data. Since the pions are the dominant
hadrons produced in $d$+Au collisions, the $Y_{\mathrm {Asym}}$  for 
$\pi^{+}$+$\pi^{-}$+$p$+$\bar{p}$ is dominated by them. 
In the absence of predictions from this model
for identified hadrons in $d$+Au data, it is not clear 
if it can describe the $Y_{\mathrm {Asym}}$ for $p$+$\bar{p}$.
One of the reasons for the success of the recombination model 
in the intermediate $p_{\mathrm T}$ range is that in this model
a baryon  is formed by
recombination of three shower and thermal partons, while a
meson needs only two, resulting in a higher yield at
larger momentum for baryons~\cite{hwa}. 
A comparison of model calculations separately for charged pions 
and $p$+$\bar{p}$ is of interest to see if  the observed weak species 
dependence and almost similar $p_{\mathrm T}$  dependence of 
$Y_{\mathrm {Asym}}$ is predicted.

\subsection{Comparison to the saturation model}
\begin{figure}
\begin{center}
\includegraphics[scale=0.45]{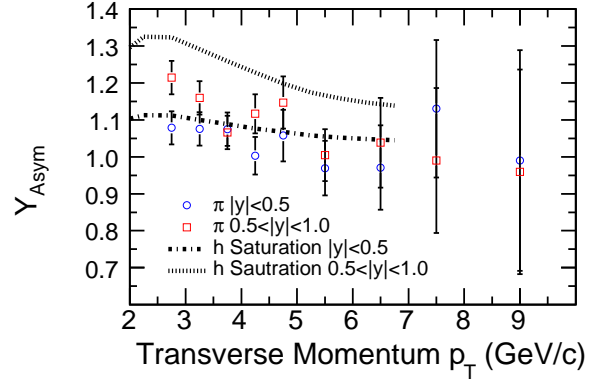}
\caption{(Color online)
High transverse momentum rapidity 
asymmetry factor ($Y_{\mathrm {Asym}}$) for 
$\pi^{+}$+$\pi^{-}$ at $\mid$$y$$\mid$ $<$ 0.5 
and 0.5 $<$ $\mid$$y$$\mid$ $<$ 1.0 for minimum bias $d$+Au collisions at 
$\sqrt{s_{\mathrm {NN}}}$ = 200 GeV compared to 
the saturation model~\cite{cgc}. See text for more details.
}
\label{fig8}
\end{center}
\end{figure}

Finally, we compare our charged pion measurements to calculations
from saturation models~\cite{cgc}. In such models the particle production is
determined by the high gluon density in the Au nucleus and the deuteron.
The model had successfully described the suppression of 
high $p_{\mathrm T}$ hadron yields at forward rapidities for 
$d$+Au data relative to $p$+$p$ data at RHIC. 
In contrast to a naive
multiple scattering picture, where one expects enhancement due to 
the Cronin effect to be more significant for larger forward 
rapidities due to the increase in the number of
scattering centers while probing smaller values of $x$, the 
saturation models give a completely opposite result~\cite{cgc1}.
For this model the momentum range 
where $Y_{\mathrm {Asym}}$~$>$~1 is determined by the
saturation and geometrical scales in the model, as well as the  
onset of the gluon saturation. 

In Fig.~\ref{fig8} we compare the $Y_{\mathrm {Asym}}$ data for charged pions 
with the $Y_{\mathrm {Asym}}$ predictions for inclusive hadrons.
Such a comparison is 
reasonable as $\pi^{+}$+$\pi^{-}$ are the dominant hadrons produced in 
$d$+Au collisions.
Further, the $Y_{\mathrm {Asym}}$ values for $\pi^{+}$+$\pi^{-}$
are similar to those for 
 $p$+$\bar{p}$. 
The model calculations are in reasonable agreement for 
$\mid$$y$$\mid$ $<$ 0.5 and give the correct decreasing trend for 
$Y_{\mathrm {Asym}}$ vs. $p_{\mathrm T}$. The prediction of a 
strong centrality dependence at midrapidity is not observed~\cite{star_asym}.
Such models are expected to work better at forward rapidities
at RHIC.  The models give larger asymmetries than 
data for 0.5 $<$ $\mid$$y$$\mid$ $<$ 1.0.

\section{NUCLEAR MODIFICATION FACTOR}

\begin{figure}
\begin{center}
\includegraphics[scale=0.45]{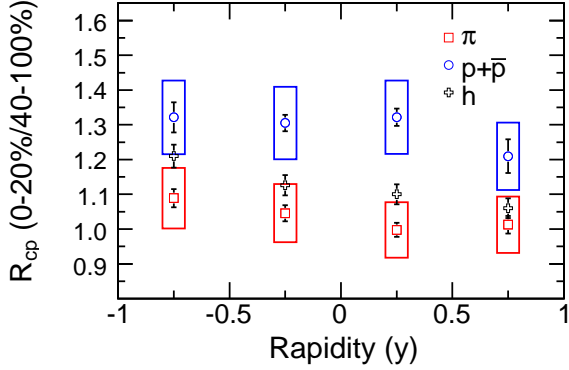}
\caption{(Color online)
Variation of nuclear modification factor ($R_{\mathrm {CP}}$) 
for $\pi^{+}$+$\pi^{-}$ and $p$+$\bar{p}$ with rapidity
for $p_{\mathrm T}$~$>$~2.5~GeV/$c$ 
for $d$+Au collisions at $\sqrt{s_{\mathrm {NN}}}$ = 200 GeV.
Also shown for comparison are the $R_{\mathrm {CP}}$ values
for inclusive charged hadrons as a function of 
pseudorapidity~\cite{star_asym}.
The errors shown as boxes are the systematic errors.
The error due to number of binary collisions is $\sim$ 14\% and 
is not shown in the figure.
}
\label{fig9}
\end{center}
\end{figure}
The gluon saturation effects are believed to manifest themselves
in terms of suppression of transverse distributions below
the saturation scale. The onset of gluon saturation and the
saturation scale, in turn, depend upon the gluon density and the
rapidity of the measured particles. The saturation scale at RHIC 
is expected to be $\sim$ 2 GeV$^{2}$ and depends on the 
colliding nuclei and rapidity as $\sim$ $A^{1/3} e^{\lambda y}$
~\cite{cgc,cgc1,cgc2}.
The value of $\lambda$ lies between 0.2--0.3 and is obtained
from fits to HERA data~\cite{hera}. It is important to study the variation
of the nuclear modification factor ($R_{\mathrm {CP}}$) as a function
of rapidity. The $R_{\mathrm {CP}}$($y$) and the 
$Y_{\mathrm {Asym}}$($p_{\mathrm T}$) together can provide a more
stringent constrain on particle production models.
Although the present data do not have large 
rapidity span, we will explore the variation of
$R_{\mathrm {CP}}$ for identified hadrons from forward to 
backward rapidity. $R_{\mathrm {CP}}$ is defined as
\begin{displaymath}
R_{CP} =
\frac{(d^{2}N/dp_{T}d\eta/\langle
N_{\rm{bin}} \rangle)|_{\rm{central}}}{(d^{2}N/dp_{T}d\eta/\langle N_{\rm{bin}} \rangle)|_{\rm{periph}}},
\end{displaymath}  
where $d^{2}N/dp_{T}d\eta$ is the differential yield per event in
$d$+Au collisions for a given centrality class.

Fig.~\ref{fig9} shows $R_{\mathrm {CP}}$ for $\pi^{+}$+$\pi^{-}$
and $p$+$\bar{p}$ with rapidity  $\mid$$y$$\mid$ $<$ 1.0.
There may be a decrease in $R_{\mathrm {CP}}$ for $\pi^{+}$+$\pi^{-}$ from
backward rapidity (gold-side) to forward rapidity (deuteron-side).
Within the systematic errors (shown as boxes) the $R_{\mathrm {CP}}$
for proton+anti-proton is almost constant within the rapidity range studied.
Also shown for comparison are the $R_{\mathrm {CP}}$ values
for inclusive charged hadrons as a function of 
pseudorapidity~\cite{star_asym}.
The dependences are slightly weaker than 
observed by BRAHMS for inclusive charged hadrons in
the forward rapidity region~\cite{brahms}.

\section{PARTICLE RATIOS}
\begin{figure}
\begin{center}
\includegraphics*[scale=0.4]{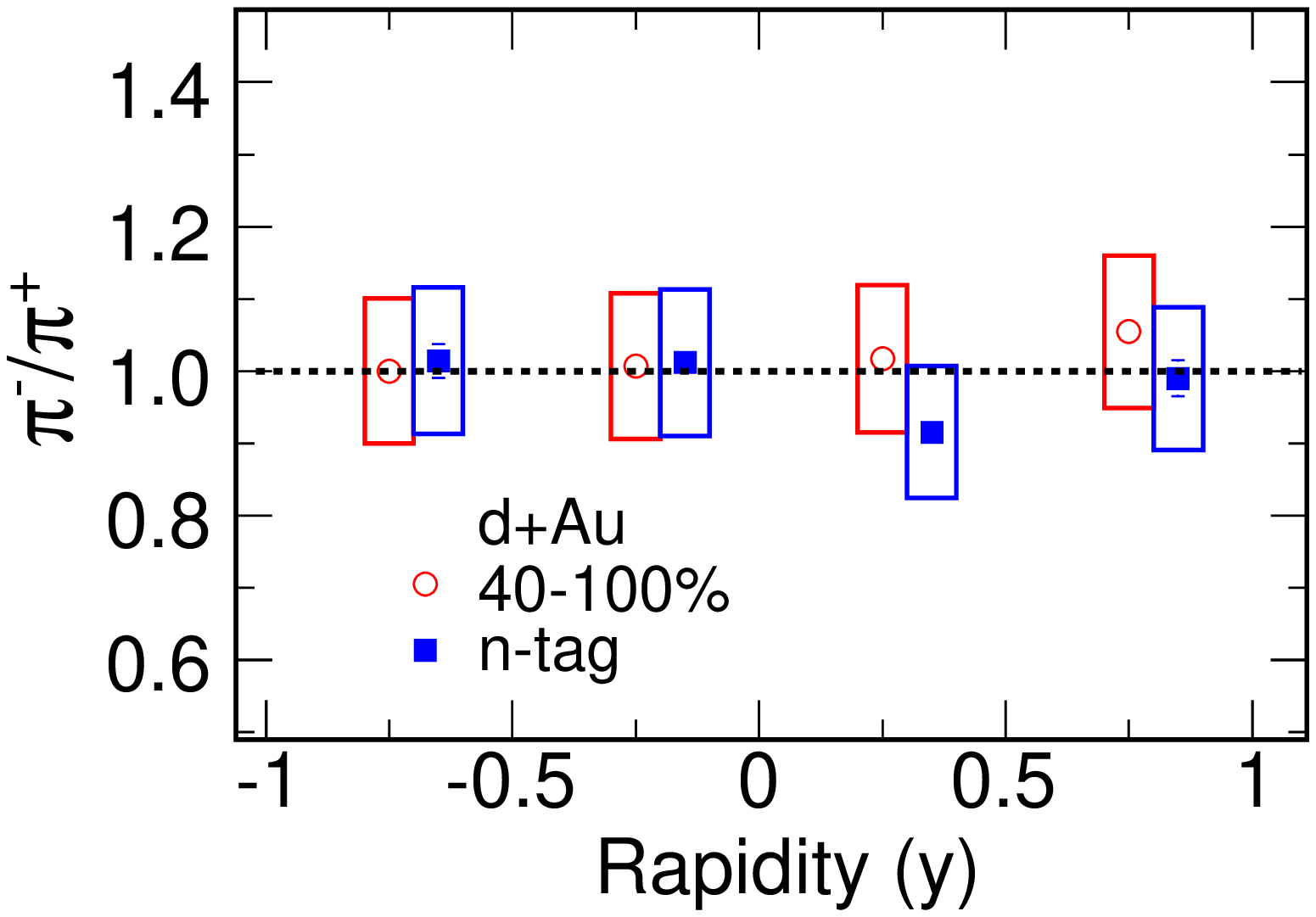}
\includegraphics*[scale=0.4]{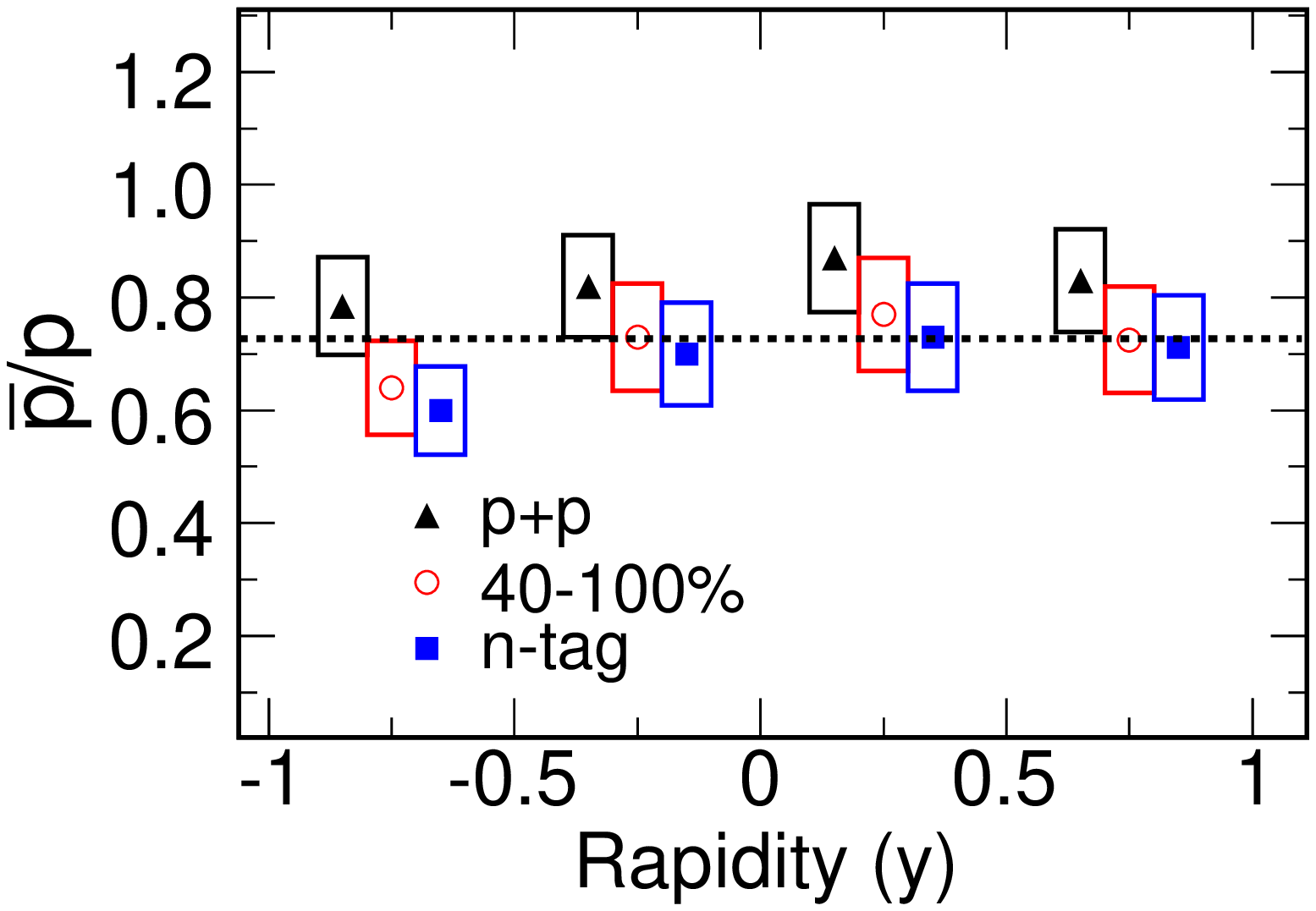}
\caption{(Color online)
Variation of $\pi^{-}$/$\pi^{+}$ and $\bar{p}$/$p$ with rapidity
for 2.5~$<$~$p_{\mathrm T}$~$<$~10~GeV/$c$ 
for peripheral (40-100\%) and $n$-tag events for 
$d$+Au collisions at $\sqrt{s_{\mathrm {NN}}}$ = 200 GeV.
Also shown for comparison are the $\bar{p}$/$p$ for minimum bias
$p$+$p$ collisions. The boxes are the systematic errors.
The ratios for $n$-tag events are shifted by 0.05 units in rapidity 
and those for $p$+$p$ collisions by -0.05 units in rapidity for 
clarity of presentation.}
\label{fig10}
\end{center}
\end{figure}

Figure~\ref{fig10} shows the $\pi^{-}$/$\pi^{+}$ and $\bar{p}$/$p$ 
ratios for 2.5~$<$~$p_{\mathrm T}$~$<$~10 GeV/$c$
as  functions of rapidity for peripheral (40-100\%) and 
$n$-tag events for $d$+Au collisions at $\sqrt{s_{\mathrm {NN}}}$ = 200 GeV.
The $\pi^{-}$/$\pi^{+}$ ratio is unity for both $n$-tag and peripheral 
events in the negative (gold-side) 
rapidity region. For the positive rapidity region, 
the absolute value of $\pi^{-}$/$\pi^{+}$ ratio is
smaller for $n$-tag events compared to 
peripheral $d$+Au data; however considering the systematic errors
(boxes), they are also consistent with unity. 
The systematic errors do not allow 
for any strong conclusions regarding
the differences, which are expected from the valence quark and 
isospin effects at high $p_{\mathrm T}$ for $n$-tag events.
The $\bar{p}$/$p$ ratios are similar within the systematic errors 
for the two event classes. The $\bar{p}$/$p$ ratios are slightly smaller 
than observed for $p$+$p$ data.

\begin{figure}
\begin{center}
\includegraphics*[scale=0.4]{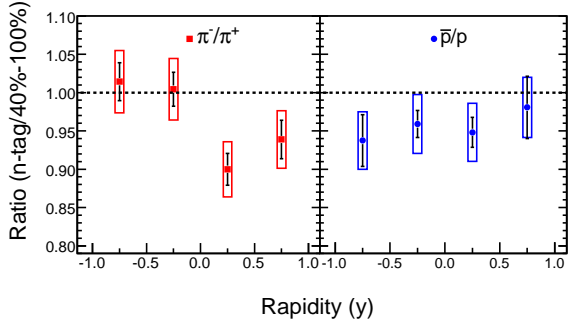}
\caption{(Color online)
Variation of double ratio 
($\pi^{-}$/$\pi^{+}$)$_{n-tag}$/($\pi^{-}$/$\pi^{+}$)$_{40-100\%}$
and ($\bar{p}$/$p$)$_{n-tag}$/($\bar{p}$/$p$)$_{40-100\%}$
with rapidity for 2.5~$<$~$p_{\mathrm T}$~$<$~10 GeV/$c$ 
for $d$+Au collisions at $\sqrt{s_{\mathrm {NN}}}$ = 200 GeV.
The boxes are the systematic errors.}
\label{fig11}
\end{center}
\end{figure}

In order to cancel out most of the systematic errors 
(listed in Table~\ref{table3}), 
we have plotted the double ratio of 
($\pi^{-}$/$\pi^{+}$)$_{n-tag}$/($\pi^{-}$/$\pi^{+}$)$_{40-100\%}$
and ($\bar{p}$/$p$)$_{n-tag}$/($\bar{p}$/$p$)$_{40-100\%}$ in
Fig.~\ref{fig11}. The double ratio clearly shows the difference
between the $\pi^{-}$/$\pi^{+}$ ratio in the forward and backward
rapidity regions when we compare peripheral $d$+Au collisions 
and $n$-tag events.
The difference for $\bar{p}$/$p$ is observed also for both rapidities.
The boxes shown in the Fig.~\ref{fig11} 
are systematic errors on the double ratio, which were 
calculated by varying: the distance of closest approach of the tracks 
from the vertex (error of $\sim$1\%), $dE/dx$ cuts 
(error of $\sim$1\%), $p_{\mathrm T}$ cuts (error of $\sim$2\%)
and  small change in rapidity range (error of $\sim$3\%). The 
total systematic error on the double ratio is $\sim$ 4\%.

\begin{figure}
\begin{center}
\includegraphics*[scale=0.4]{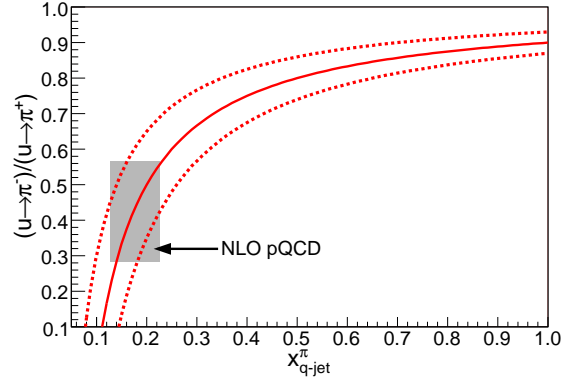}
\caption{(Color online)
Variation of 
$\frac{u\rightarrow \pi^{-} }{u\rightarrow \pi^{+}}$
with fraction of pions ($x^{\pi}_{q-jet}$)
originating from quark jets (solid line) obtained from the 
measured ratios of $\pi^{-}$/$\pi^{+}$ in peripheral $d$+Au collisions
and $n$-tag events. 
The dashed lines reflects 1$\sigma$ systematic 
errors in obtaining the 
$\frac{u\rightarrow \pi^{-} }{u\rightarrow \pi^{+}}$.
The horizontal shaded band reflects the possible $x^{\pi}_{q-jet}$
range obtained from NLO pQCD calculations using AKK fragmentation
functions~\cite{akk}.}
\label{fig12}
\end{center}
\end{figure}

The particle ratios and the double ratios can be used to
get an idea of relative fragmentation of $d$ and $u$ quarks
to protons, as well as $u$-quarks to $\pi^{+}$ and $\pi^{-}$.
Assuming that gluons fragment equally to $\pi^{+}$ and $\pi^{-}$ 
at high $p_{\mathrm T}$ and using the measured double ratio of  
$\pi^{-}$/$\pi^{+}$ (Fig.~\ref{fig11}), one can show that the
variation of the ratio of $u$-quark fragmenting to $\pi^{-}$
to $u$-quark fragmenting to $\pi^{+}$ 
is sensitive to the fraction of pions ($x^{\pi}_{q-jet}$)
originating from quark jets: 
\begin{displaymath}
\frac{u\rightarrow \pi^{-} }{u\rightarrow \pi^{+}} = 
1-\frac{1-r_{\pi}}{x^{\pi}_{q-jet}}  ,
\end{displaymath}  
where $r_{\pi}$ is the double ratio of $\pi^{-}$/$\pi^{+}$
shown in Fig.~\ref{fig11}. The results are shown in Fig.~\ref{fig12}.
The dashed lines reflect the 1$\sigma$ uncertainty in the ratio
$\frac{u\rightarrow \pi^{-} }{u\rightarrow \pi^{+}}$ due to uncertainty
in the measurements of the double ratio of $\pi^{-}$/$\pi^{+}$. 
The horizontal shaded band reflects the $x^{\pi}_{q-jet}$ value 
for charged pions from NLO pQCD calculations using the Albino-Kniehl-Kramer
(AKK) set of fragmentation functions (FFs)~\cite{akk}. The width of 
this band reflects the uncertainty
associated with $x^{\pi}_{q-jet}$ from the NLO pQCD calculations. 
These are obtained by varying the factorization scales from 
0.5$p_{\mathrm T}$ to 2$p_{\mathrm T}$. Since 
the NLO pQCD calculations with AKK FFs
agree reasonably well with charged pion measurements at RHIC, from the
figure we can conclude that 
$\frac{u\rightarrow \pi^{-} }{u\rightarrow \pi^{+}}$ $\sim$ 0.3 to 0.6.

Similarly, it can be shown that the ratio of $u$-quark fragmenting 
to protons, to $d$-quark fragmenting to protons is given as,
\begin{displaymath}
\frac{u\rightarrow p}{d\rightarrow p} = 
1+\frac{(1-r_{p2})}{(1-r_{p1})} \sim 1.3\pm 0.42  ,
\end{displaymath}  
where $r_{p1}$ is the ratio $\bar{p}$/$p$ $\sim$ 0.81, 
and $r_{p2}$ is the double 
ratio of $\bar{p}$/$p$ shown in Fig.~\ref{fig11}.
The above relation is valid for the assumption that all the 
anti-protons at high $p_{\mathrm T}$ are from 
gluon fragmentation and that gluons fragment equally to protons 
and anti-protons. Similar measurements have been carried out by
the OPAL collaboration for $e^{+}+e^{-}$ collisions as a function of 
$x_{p}$ (2p/$\sqrt{s}$, largest scaled momentum)~\cite{opal}. 
They observed the above ratio to be 
0.637~$\pm$~0.173~$\pm$~0.083 for $x_{p}$~$>$~0.2 
and that it decreases to 0.165~$\pm$~0.421~$\pm$~0.257 
for $x_{p}$~$>$~0.5.

\section{Summary}
We have presented  transverse momentum spectra for
identified charged pions, protons and anti-protons from $d$+Au
collisions in various centrality classes at $\sqrt{s_{\mathrm {NN}}}$
= 200 GeV. The transverse momentum spectra are measured in
4 rapidity bins for $-1<y<1$ over the range 
2.5~$<$~$p_{\mathrm T}$~$<$~10 GeV/$c$. The rapidity, $p_{\mathrm T}$,
centrality, and species dependence of the rapidity asymmetry $Y_{\mathrm {Asym}}$
has been studied.
We have also presented the rapidity dependence of the nuclear
modification factor and the $\pi^{-}$/$\pi^{+}$ and $\bar{p}$/$p$ ratios
for the rapidity range $\mid$$y$$\mid$~$<$~1.0 
and $p_{\mathrm T}$~$>$~2.5~GeV/$c$.

The $Y_{\mathrm {Asym}}$ is found to be larger for
0.5~$<$ $\mid$$y$$\mid$~$<$~1.0 than for
$\mid$$y$$\mid$~$<$~0.5 in the range 2.5~$<$~$p_{\mathrm T}$~$<$~5.0~GeV/$c$. 
For higher $p_{\mathrm T}$ the $Y_{\mathrm {Asym}}$ 
approach 1 for
both charged pions and $p$+$\bar{p}$. From these
observations we conclude that possible sources of
nuclear effects in $d$+Au collisions, such as parton
saturation, nuclear shadowing, or energy loss in cold nuclear
matter, have a strong rapidity dependence which vanishes for
$p_{\mathrm T}$~$>$~5.5 GeV/$c$. The observed
$Y_{\rm{Asym}}$ vs. $p_{\mathrm T}$ dependence rules out
models based on incoherent initial multiple partonic scattering
and independent fragmentation.

Comparison to models based on nuclear shadowing reveals that
incorporation of extremes of gluon shadowing at low $x$ does
not reproduce the measured $Y_{\rm{Asym}}$. This provides  
an upper limit on the contribution of nuclear shadowing to the $Y_{\rm{Asym}}$.
Models incorporating multiple scattering, dynamical shadowing, and
energy loss in cold nuclear matter are in reasonable agreement
with the data for $\mid$$y$~$\mid$~$<$ 0.5. However, the 
$Y_{\rm{Asym}}$ being independent of $p_{\mathrm T}$ ($>$ 3 GeV/$c$) 
is inconsistent with the measurements at higher rapidity.  
Qualitatively, features of
monotonic decrease in $Y_{\rm{Asym}}$ with $p_{\mathrm T}$ and
$R_{\mathrm {CP}}$ with $y$ are in agreement with color-glass-condensate (CGC) type models.
However, there is a lack of quantitative agreement at higher rapidities
where this model is expected to work better.
Further, the absence of very strong centrality dependence at midrapidity
in the data is in contrast to the predictions from CGC models.

The EPOS and recombination
models are in best quantitative agreement with the data. The actual test
of the recombination model is only possible when the calculations are available
for $Y_{\rm{Asym}}$ for identified baryons and mesons. 
It will be interesting to see if this model can explain the observed 
weak species dependence and similar $p_{\mathrm T}$ dependence of 
$Y_{\mathrm {Asym}}$ for $\pi^{+}$+$\pi^{-}$ and $p$+$\bar{p}$.

In general, the study of identified hadron $Y_{\mathrm
{Asym}}$ as a function of many variables ($y$, $p_{\mathrm T}$, centrality and
particle type) for $d$+Au collisions has been able to provide some
definitive insight on mechanisms of particle production in $d$+Au
collisions at  $\sqrt{s_{\mathrm {NN}}}$ = 200 GeV.
The $Y_{\mathrm {Asym}}$($p_{\mathrm T}$) together with $R_{\mathrm {CP}}$($y$) 
can provide a more stringent constrain on particle production models.
It may be mentioned that a detailed of study of 
particle yields ($p_{\mathrm T}$ $<$ 3 GeV/$c$) at midrapidity and 
forward rapidity in STAR has revealed a possible alternative explanation 
of the pseudorapidity dependence of $R_{\mathrm {CP}}$ from a purely 
geometrical picture.  The decrease 
in $R_{\mathrm {CP}}$ from negative (backward) to positive (forward) 
rapidity can be explained by considering the inital asymmetry in 
particle production in $d$+Au collisions compared to 
the symmetric $p$+$p$ collisions~\cite{joern}.

The ratios $\pi^{-}$/$\pi^{+}$ and $\bar{p}$/$p$ have been studied
for peripheral and $n$-tag events for $d$+Au collisions
to see the possible valence quark effect.
For the range 2.5~$<$~$p_{\mathrm T}$~$<$~10~GeV/$c$ and the
rapidity region on the deuteron-side, the ratios for $n$-tag events are
smaller than for peripheral events. However, within the systematic errors
it is difficult to make strong conclusions
of valence quark effects on particle production at high
$p_{\mathrm T}$. The $\bar{p}$/$p$ ratios are observed to be
systematically lower than corresponding values from $p$+$p$
collisions.

The double ratio between $n$-tag events and 40-100\% peripheral collision
events does reveal a clear enhancement in $\pi^{+}$
production relative to $\pi^{-}$ at forward rapidity (deuteron-side).
No such enhancement is observed at the backward rapidity (gold-side).
Using the above ratio measurements we have found
$\frac{u\rightarrow \pi^{-} }{u\rightarrow \pi^{+}}$
$\sim$ 0.3 - 0.6 and
$\frac{u\rightarrow p}{d\rightarrow p}$ $\sim$ 1.3 $\pm$ 0.42
for 2.5~$<$~$p_{\mathrm T}$~$<$~10~GeV/$c$.

A future, high statistics run for $d$+Au collisions 
at RHIC may be able to provide data that will lead to a still better insight into valence
quark and gluon contribution, as well as isospin effects at high
$p_{\mathrm T}$.

\normalsize

\begin{acknowledgments}        
We would like to thank S. Albino, R. Hwa, 
I. Vitev, R. Vogt, K. Werner, and C. B. Yang
for providing us the results for the different model calculations
and many useful discussions.
We thank the RHIC Operations Group and RCF at BNL, and the
NERSC Center at LBNL for their support. This work was supported
in part by the Offices of NP and HEP within the U.S. DOE Office 
of Science; the U.S. NSF; the BMBF of Germany; CNRS/IN2P3, RA, RPL, and
EMN of France; EPSRC of the United Kingdom; FAPESP of Brazil;
the Russian Ministry of Science and Technology; the Ministry of
Education and the NNSFC of China; IRP and GA of the Czech Republic,
FOM of the Netherlands, DAE, DST, and CSIR of the Government
of India; Swiss NSF; the Polish State Committee for Scientific 
Research; SRDA of Slovakia, and the Korea Sci. \& Eng. Foundation.
\end{acknowledgments}


\end{document}